\documentclass{article}
\usepackage{spconf,amsmath,graphicx,booktabs}

\title{Synthesis and Edition of Ultrasound Images via Sketch Guided \\Progressive Growing GANs}
\name{\parbox{\linewidth}{\centering Jiamin Liang$^{1,2,\dagger}$, Xin Yang$^{1,2,\dagger}$, Haoming Li$^{1,2}$, Yi Wang$^{1,2}$, Manh The Van$^{1,2}$, Haoran Dou$^{1,2}$, \textit{Chaoyu Chen$^{1,2}$, Jinghui Fang$^{3}$, Xiaowen Liang$^{3}$, Zixin Mai$^{3}$, Guowen Zhu$^{1,2}$, Zhiyi Chen$^{3}$, Dong Ni$^{1,2,*}$}} \thanks{$\dagger$ Authors contributed equally.} \thanks{* Corresponding author: \textit{nidong@szu.edu.cn}.} \thanks{The work in this paper was supported by the grant from National Key R\&D Program of China (No. 2019YFC0118300); Shenzhen Peacock Plan (No. KQTD2016053112051497, KQJSCX20180328095606003); Medical Scientific Research Foundation of Guangdong Province, China (No. B2018031).}}

\address{$^{1}$National-Regional Key Technology Engineering Laboratory for Medical Ultrasound, \\Guangdong Key Laboratory for Biomedical Measurements and Ultrasound Imaging, \\School of Biomedical Engineering, Health Science Center, Shenzhen University, Shenzhen, China \\$^{2}$Medical UltraSound Image Computing (MUSIC) Lab, Shenzhen University, Shenzhen, China \\$^{3}$Experimental Center of Liwan Hospital, The Third Affiliated Hospital of \\Guangzhou Medical University, Guangzhou, China.}

\begin{document}
\maketitle

\begin{abstract}
Ultrasound (US) is widely accepted in clinic for anatomical structure inspection. However, lacking in resources to practice US scan, novices often struggle to learn the operation skills. Also, in the deep learning era, automated US image analysis is limited by the lack of annotated samples. Efficiently synthesizing realistic, editable and high resolution US images can solve the problems. The task is challenging and previous methods can only partially complete it. In this paper, we devise a new framework for US image synthesis. Particularly, we firstly adopt a sketch generative adversarial networks (\textit{Sgan}) to introduce background sketch upon object mask in a conditioned generative adversarial network. With enriched sketch cues, Sgan can generate realistic US images with editable and fine-grained structure details. Although effective, Sgan is hard to generate high resolution US images. To achieve this, we further implant the Sgan into a progressive growing scheme (\textit{PGSgan}). By smoothly growing both generator and discriminator, PGSgan can gradually synthesize US images from low to high resolution. By synthesizing ovary and follicle US images, our extensive perceptual evaluation, user study and segmentation results prove the promising efficacy and efficiency of the proposed PGSgan. \par
\end{abstract}


\section{Introduction}
Featured as real-time and radiation-free, ultrasound (US) is preferred in clinic for anatomical structure inspection. US-based diagnoses require sonographers to have strong skills in scanning subject and interpreting US images. However, due to the lack of clinic resources and opportunities to practice with US machines, novices often need to struggle for a long time to obtain the required skills. Realistic US image-based scan training is hence highly desired \cite{simulation2004}. On the other hand, since labeling US images requires expertise and is time-consuming, the lack of large-scale annotated US dataset severely confines the development of automated US image analyses, especially in the deep learning era \cite{liu2019deep}. Regarding this situation, synthesizing US images has great potentials in facilitating both the clinical training and technical evolution. \par

\begin{figure}
	\centering
	\includegraphics[width=0.95\linewidth]{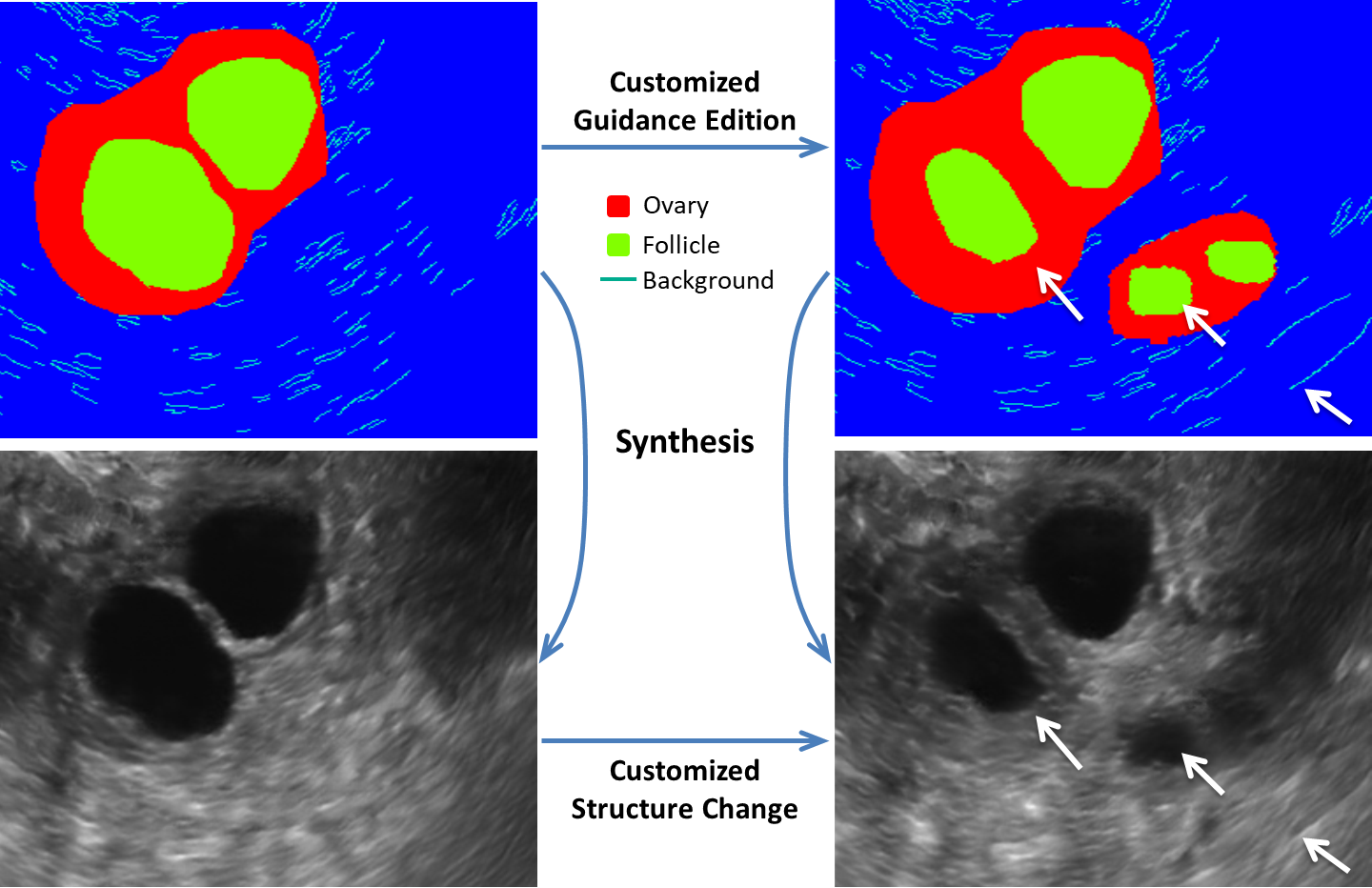}
	\label{fig:problem-task}
	\caption{Three changes are edited in Canny label to synthesize a new ultrasound images. The white arrow marks the place edited. In order to make it clearer, we present Canny sketch in pseudo color.}
	\vspace{-0.4cm}
\end{figure}

US image synthesis should be realistic with high resolution to provide an immersive experience, be editable to mimic various disease cases and be efficient for clinic deployment. Previous researches have made continuous efforts but only partially completed these criteria.
Traditional physical principle based methods, such as kidney phantom images synthesis in Field II \cite{fieldii}, the intravascular US (IVUS) images synthesis \cite{simulation2004} and cardiac images sequence generation \cite{cardiacgeneration2012,zhou2017framework}, though containing clearer physical parameters for synthesis, they have high computational complexity and are time-consuming when generating a new image, especially for high-resolution images. 
In recent years, using deep neural network based generator and discriminator to play a min-max game, Generative Adversarial Networks (GANs) \cite{GAN} provide a brand new way to efficiently synthesize realistic images. Once the training is done, it takes only a few seconds for a new image generation. However, the random noise input of GAN is not editable and thus can not be customized in detail. Conditional GAN (cGAN) \cite{cGAN} enables the user-controlled image generation. Tom et al. \cite{simulating2018} introduced a multi-stage method for IVUS synthesis, including two different cGANs to transform tissue maps into IVUS images. Due to the cascaded connection between two different cGANs, the system is hard to train. In \cite{freehand2017}, spatially-conditioned GAN was proposed to synthesize US images from fetal phantom. Although effective, the synthesized US images are with low-resolution and possible checkerboard artifacts. To enrich the structure details for easy training of cGAN and more realistic synthesis, the auxiliary guidance information, like the sketch and edge of background were introduced in \cite{shin2018abnormal,zhang2019skrgan}. However, these methods still have problems in generating high-resolution images. Generating high-resolution images is challenging because higher resolution makes it easier for the discriminator to capture flaws in the synthesized images. Also, higher resolution restricts the use of large minibatches due to memory constraints, which further degrades the training stability. \par

In this paper, we devise a new end-to-end framework to efficiently synthesize realistic, high-resolution and editable US images. In order to enhance the structure fidelity and ease the synthesis, we firstly adopt a \textit{Sgan} to introduce auxiliary sketch guidance upon object mask in a cGAN (Fig.~\ref{fig:problem-task}). Sgan enables the interactive edition of fine-grained details. Although effective, Sgan is hard to generate high resolution US images. To achieve this and avoid the use of multiple cGANs, we apply a progressive training strategy \cite{PGGAN} on Sgan to gradually generate high-resolution US images (\textit{PGSgan}). By smoothly growing both the generator and discriminator with fade-in blocks (\textit{FIB}) for transition, PGSgan can successfully learn to synthesize high resolution images with less training time. We validated PGSgan on ovary and follicle US image synthesis. Extensive perceptual evaluation, customized edition, user studies and segmentation tests prove the promising efficacy and efficiency of PGSgan. Our synthesized 512$\times$512 images may promote the clinic training and automated analysis. \par

\section{Methodology}
Fig. \ref{fig:Frame-work} shows the workflow of our proposed PGSgan. The system takes label and sketch guidance as input, and outputs realistic US images from low to high resolutions. Only one generator and discriminator in the framework. Under our progressive learning strategy, generator and discriminator are growing during training process, with all layers in both networks being learnable throughout the training process. \par

\subsection{Backbone of Sgan}
\label{sec:backbone}
\begin{figure}[t]
	\centering
	\includegraphics[width=1.0\linewidth]{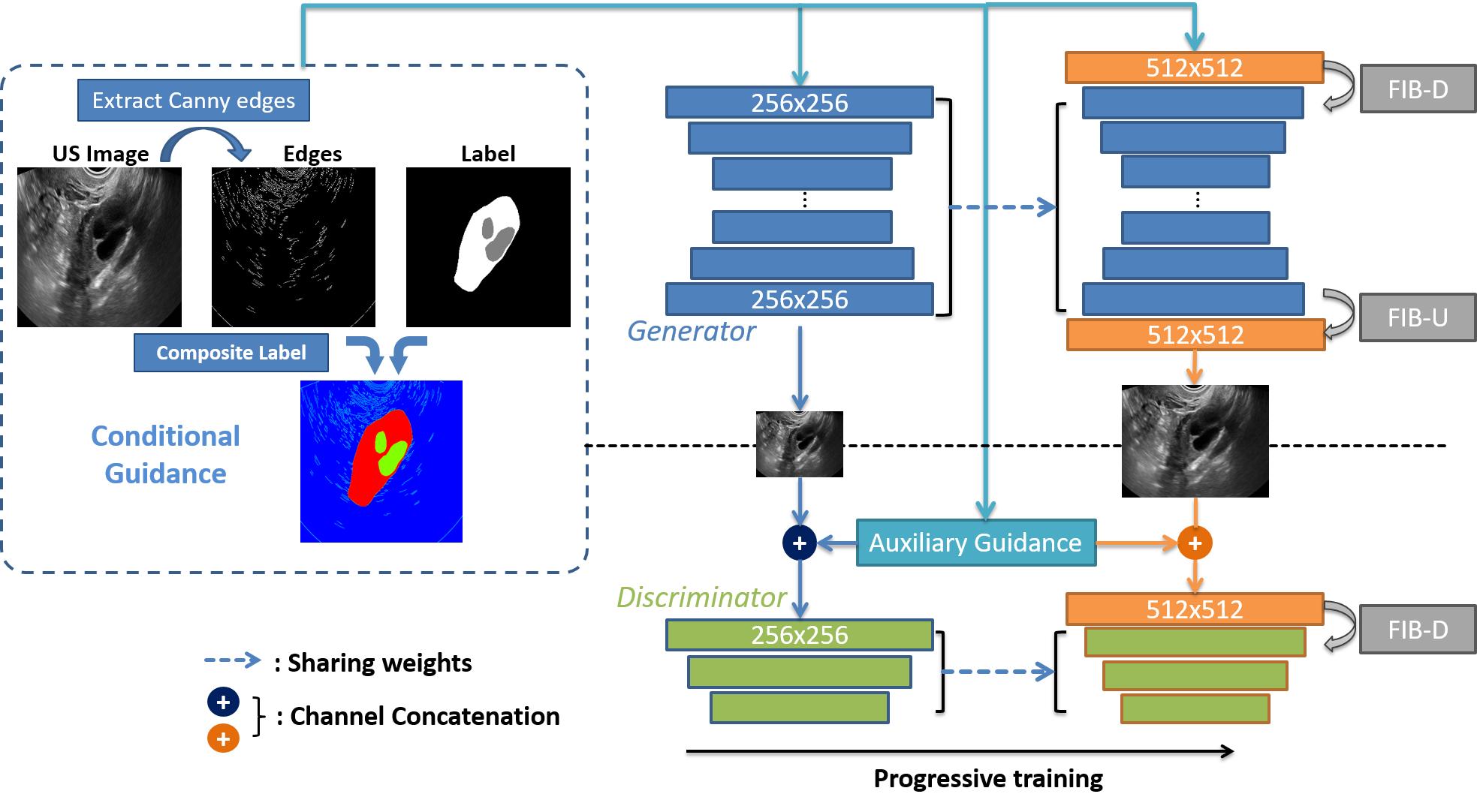}
	\vspace{-0.3cm}
	\caption{PGSgan framework. Generator input is label mask and edge sketch. Input resolution increases as the generator and discriminator grow. FIB transits among high- and low-resolution layers. FIB-D for downsampling, FIB-U for upsampling. N$\times$N is resolution. Yellow box is layer growing.}
	\label{fig:Frame-work}
\end{figure}

As shown in Fig. \ref{fig:Frame-work}, in order to balance between the training cost of Sgan and synthesis quality, we set the generator as a encoder-decoder architecture with a 10-residual block encoder. For the discriminator, we follow the design of patch-based PatchGAN \cite{cGAN} with a 30$\times$30 output, which can force the generator produce more realistic results than image-based discriminator does. With these settings, the results have better performance through experiments. PatchGAN restricts the discriminator to only model high-frequency structure details, while exploits a \textit{L1}-loss to force low-frequency synthesis. The conditional adversarial loss and \textit{L1}-loss of Sgan are formulated as follow:\par
\begin{scriptsize}
	\begin{align}
	\label{eq:hybrid}
	L_{sgan}(G,D) &= E_{x,y}[log D(x,y)] + E_{x,G(x)}[log(1 - D(x,G(x))]\\
	\label{eq:L1_loss}
	L_{L1}(G) &= \lVert y - G(x) \rVert _1\\
	\label{eq:total_loss}
	G* &= arg \min _G \max _D L_{Sgan}(G,D) + L_{L1}(G)
	\end{align}
\end{scriptsize}
G, D represent the generator and discriminator, respectively. $x$ denotes the condition image input, which will be elaborated in Sec.\ref{sec:canny-edges}. $y$ denotes the ground truth US images. The training of Sgan starts from the resolution of 256$\times$256. \par

\subsection{Auxiliary Sketch Guidance}
\label{sec:canny-edges}
For cGAN training, a pair of images, i.e., input conditional image and real image, are required. In this work, we use the segmentation mask of object to serve as the basic input conditioned image. We observed that, due to the absence of the background pattern, only using the mask as input can not get realistic and natural synthesis. \par

To alleviate the problem, motivated by \cite{shin2018abnormal}, we propose to introduce auxiliary sketch guidance of the background to form composite conditional input. As shown in Fig.\ref{fig:Frame-work}, by injecting the sketch features to represent the textures of background tissues, we upgrade the paired conditional input to a triplet version. Specifically, \textit{Canny} algorithm \cite{canny} is adopted to effectively extract binary edges as sketch, especially the weak edges. Canny edge is robust against noise and avoid redundant edges. With the edge sketch, the conditional label of our system is created by overlaying the auxiliary sketch onto the original mask without affecting the area of target object. \par

With the auxiliary sketch of background, we can generate realistic US images and avoid synthesizing unnatural images in which the background has blurred areas and distored structures (see Section \ref{sec:results}). \par

\subsection{Progressive Growing Scheme}
\label{sec:pregressve growing}
High resolution amplifies the flaws in structure details of synthesized US images and hence is difficult to achieve. To tackle the challenges and avoid heavy computation, we propose to adopt the progressive growing training scheme \cite{PGGAN} to decompose the task as an incremental learning task (PGSgan). By gradually growing the layout of both generator and discriminator for capacity enhancement, the scheme enables PGSgan to use only one generator and discriminator with fast and smooth learning for high-resolution, realistic synthesis. \par

\begin{figure}[t]
	\centering
	\includegraphics[width=1.0\linewidth]{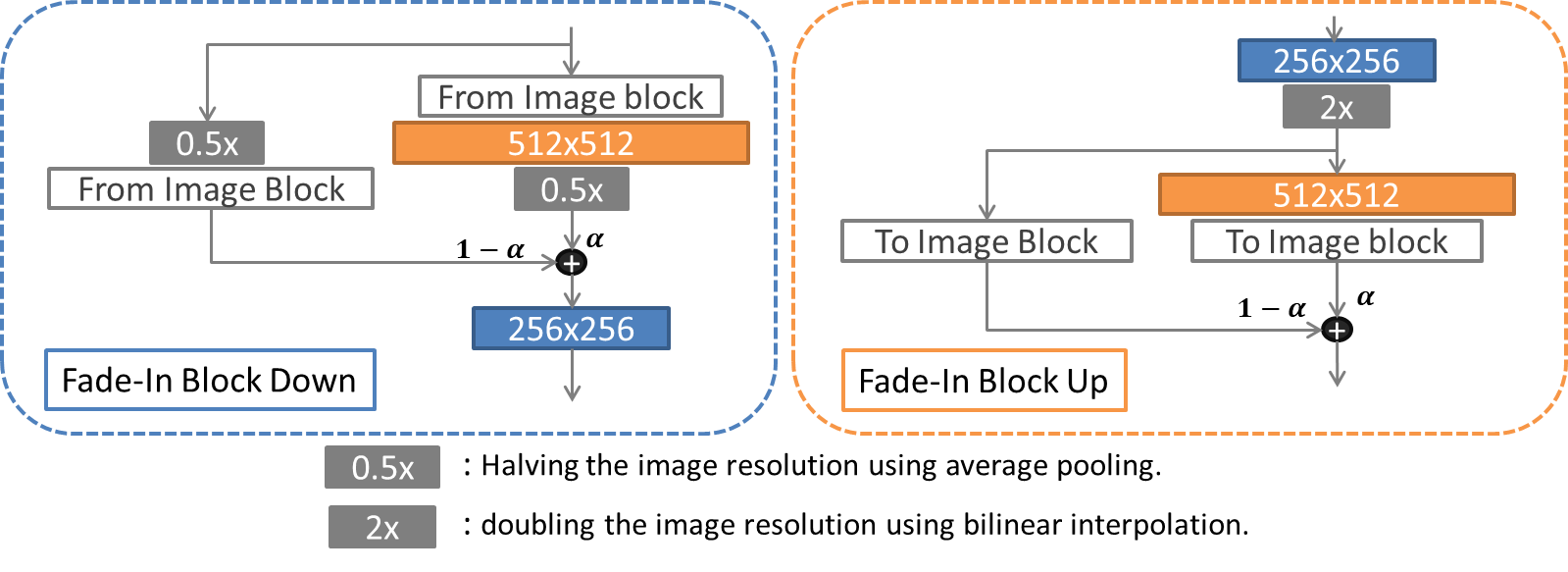}
	\caption{Structure of FIB-D (left) and FIB-U (right). Convolution in colored box. $\alpha$ increases linearly for transition. N$\times$N blocks refer to convolutional layers operating on corresponding resolution.}
	\label{fig:Fadein}
	\vspace{-0.3cm}
\end{figure}

Accordingly, we start the training of PGSgan on low-resolution conditional input (\textit{256$\times$256}). We then progressively increase the input resolution and add learnable layers to the existing network with sharing weights for continuous training (Fig.~\ref{fig:Frame-work}). To avoid the sudden shock on training when adding new layers, we adopt the fade-in block (FIB) for smooth transition in both generator and discriminator. \par

Fig. \ref{fig:Fadein} illustrates the structure details of FIB. For the purpose of transition, FIB follows the residual design. Facing with the different resolutions between the input and internal layers, FIB applies the added convolution layers and resizes the input in the main stream to match the output resolution. At the same time, FIB introduces a skip connection to directly resize the input and skips it to merge with the main stream through the weight $\alpha \sim (0,1)$. $\alpha=1$ means the output does not need the original input. During training, the transition happens as the $\alpha$ increases and the output depends less on original input resolution. The advantages of FIB are that it not only smooths the transition between different resolutions, but also remains the base model structure, making weights sharing possible and hence reducing training time. Specifically, there are two types of FIB. FIB-D is for downsampling, which is used in both generator and discriminator. FIB-U is for upsampling, which is only used in generator (Fig. \ref{fig:Frame-work}).  \par

\subsection{Training with FIB}
To ease the learning of PGSgan, we conduct the training with four phases. In the first phase, PGSgan is trained with resolution $256\times 256$ US image and composite sketch label until convergence. In the second phase, we increase the input resolution to $512\times 512$ and grow the discriminator of PGSgan by adding a FIB-D after input. Generator then grows in the third phase to match the increased input resolution. For end-to-end requirement, a FIB-D is added after the input while another FIB-U is added before the output (Fig. \ref{fig:Frame-work}). The discriminator grows earlier than generator to replenish the gradient information and force the generator learn to synthesize higher resolution. Finally, we train the PGSgan in resolution $512\times 512$ until convergence. \par

Every time the network grows for a new training phase, we increase the $\alpha$ linearly from 0 with an interval of 1/30. Notably, because we observed that allowing the $\alpha$ increases to 1 when growing generator could cause a sudden increase in generator loss, we propose to truncate the $\alpha$ at 0.5 to partially preserve the original input information for better stability. \par

\section{Experimental Results}
\label{sec:results}

Experiments were carried on a dataset of 3261 ovarian US images. All images were labeled by experienced doctors. Data collection has been approved by local IRB. We used 2848 images for training and the rest 431 images for testing. We adopted Adam with a batch size of 4, on a single TITAN X GPU with 12GB RAM, to train the whole framework. The learning rate of generator and discriminator was set to 0.001 and 0.0001, respectively.
The maximum epoch was 200.\par

\begin{figure}[t]
	\centering
	\includegraphics[width=0.98\linewidth]{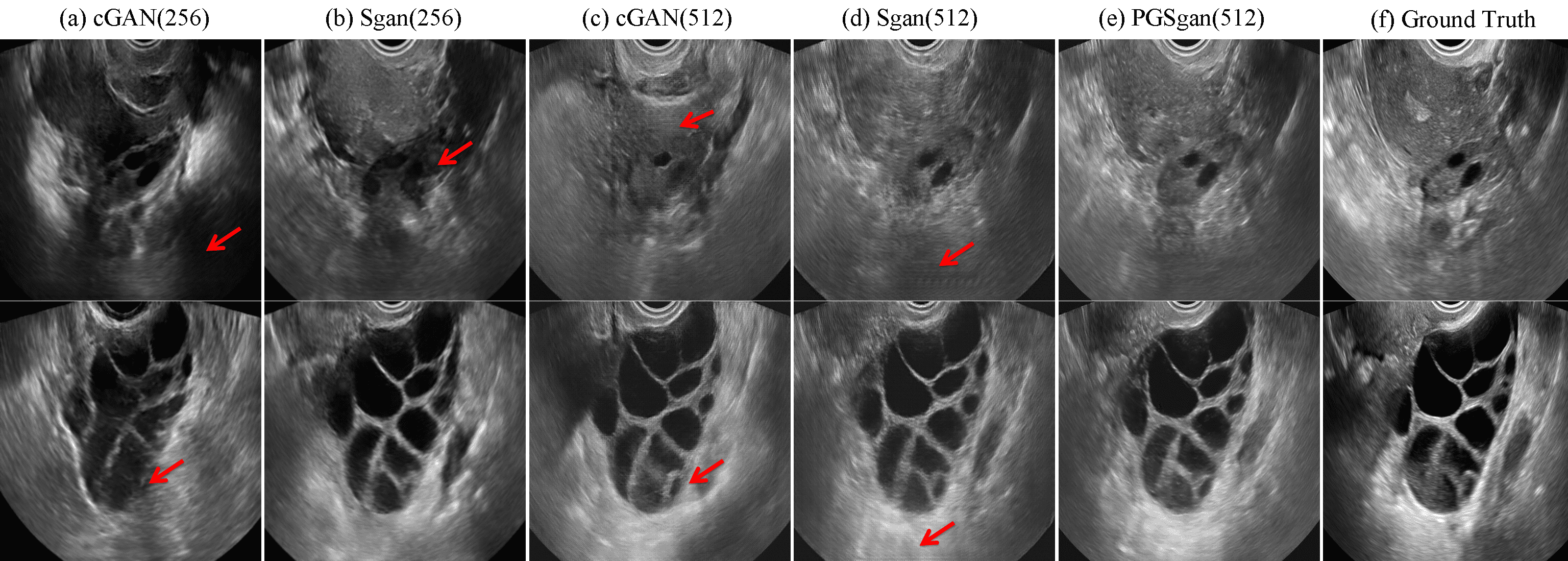}
	\vspace{-0.3cm}
	\caption{Visual comparison of some synthesized images from different methods at the resolution of 512$\times$512.}
	\label{fig:Qual-img}
\end{figure}

\begin{figure}[t]
	\centering
	\includegraphics[width=0.98\linewidth]{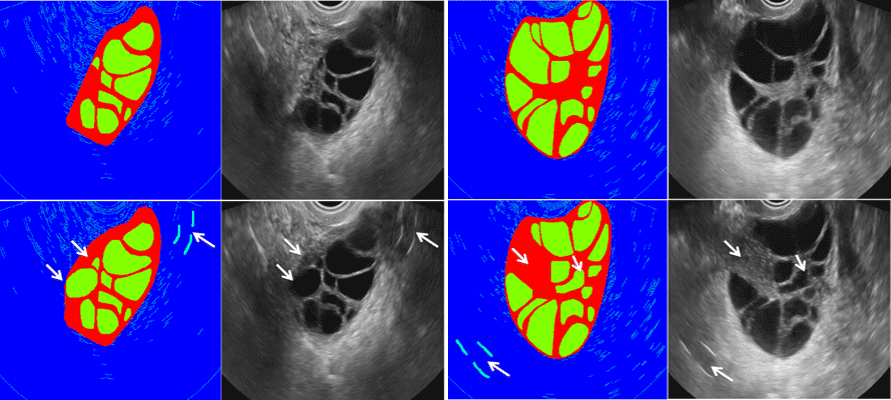}
	\vspace{-0.3cm}
	\caption{Customized synthesis of ovarian US images based on label edition. First/second row: before/after editing.}
	\label{fig:edit_show}
\end{figure}

\begin{figure}[t]
	\centering
	\includegraphics[width=1.0\linewidth]{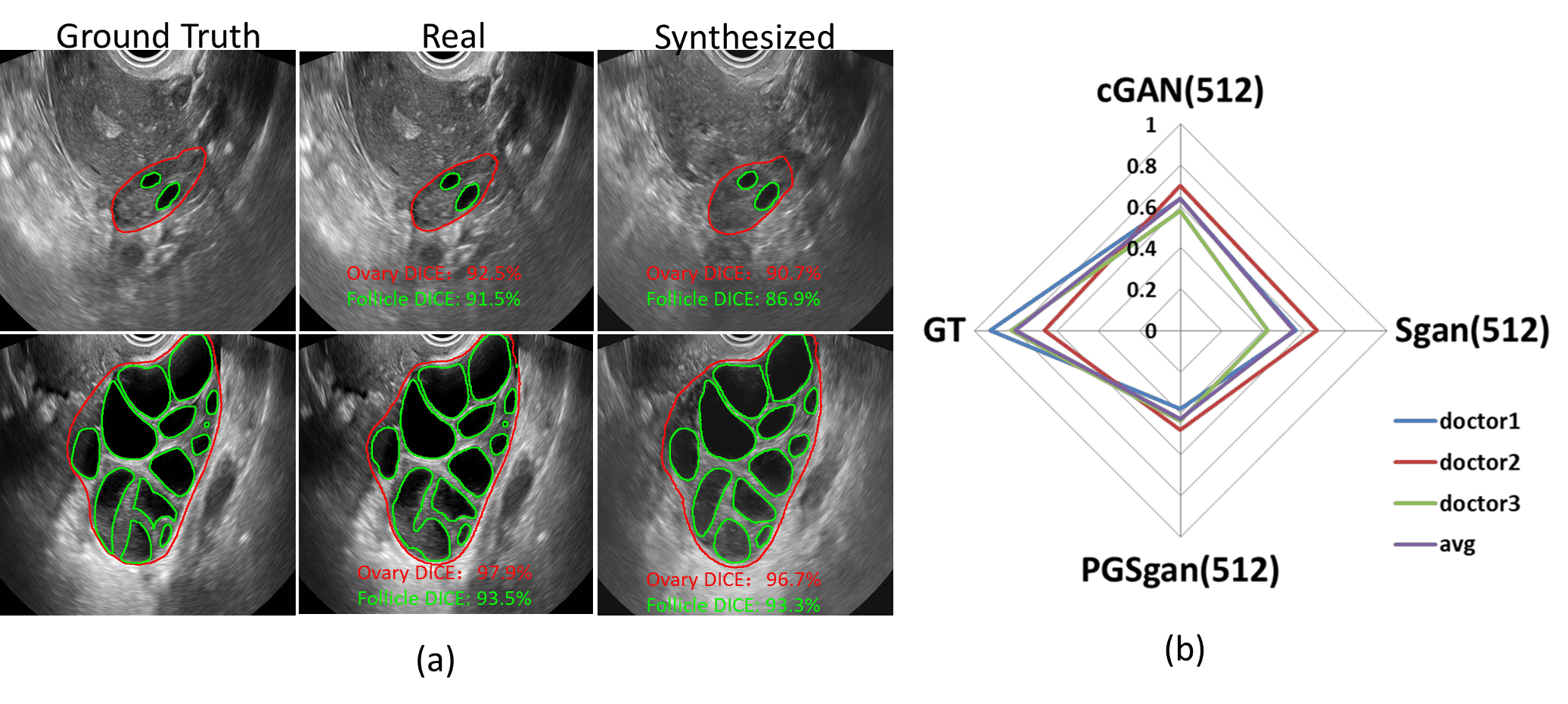}
	\vspace{-0.5cm}
	\caption{(a) Ovary and follicle segmentation results of real and synthesized images. (b) Radar chart of user study results, which shows the accuarcy of doctors' assessments.}
	\label{fig:combine_seg_user}
\end{figure}

\begin{table}[t]
	\centering
	\scriptsize
	\caption{Evaluation results for synthesized images.}\label{tab:result}
	\begin{tabular}{l|c|c|c|c}
		\toprule
		Methods& Resolution& FID$\downarrow$& KID($\times$100)$\downarrow$& MS-SSIM$\uparrow$\\
		\bottomrule
		cGAN\cite{cGAN} &256 &67.52 &8.38 &0.2404\\
		\hline
		Sgan &256 &65.95 &6.31 &0.4355\\
		\hline
		cGAN\cite{cGAN} &512 &104.95 &14.18 &0.2536\\
		\hline
		Sgan &512 &79.25 &7.96 &0.4758\\
		\hline
		PGSgan &\textbf{512}& \textbf{54.94}& \textbf{4.12}& \textbf{0.4895}\\
		\bottomrule
	\end{tabular}
\end{table}

Fig.~\ref{fig:Qual-img} visualizes the comparison of generated US images from different methods.
Some defects such as background distortion (a,b), checkerboard artifacts (c) and stretching effects (d) are denoted by arrows in Fig.~\ref{fig:Qual-img}.
In contrast, our method generates more realistic synthesized images. Furthermore, our method enables not only synthesizing images from real labels but also being capable to generate images using edited labels (see Fig.~\ref{fig:edit_show}).
By this way, we can create high fidelity ovarian US images with various pathological morphologies.
Table~\ref{tab:result} further lists the numerical evaluation metrics, including Freshet Inception Distance (FID)~\cite{FID}, Kernel Inception Distance (KID)~\cite{KID} and multi-scale structural similarity (MS-SSIM)~\cite{MS-SSIM}. 8.5 hours was needed for training the resolution of 256 and 16 hours for resolution of 512. For testing phase, it took only 0.07s for an image generation in GPU and 5.58s in CPU on average.
Our method outperformes all other compared methods with regard to all metrics, which is mainly due to the employment of auxiliary sketch and progressive training scheme.\par

We further adopted U-Net\cite{ronneberger2015u} to investigate the segmentation performance on synthesized images. The average dice of ovary and follicle are 0.92 and 0.89, respectively. Fig.~\ref{fig:combine_seg_user} (a) shows that the segmentation on synthesized images are comparable with results from real images, which proves our synthesis results could benefit the development of automatic segmentation through realistic data augmentation. \par

To further validate the quality of synthesized images, a user study of blinded trials was designed with three doctor participants.
They tried to tell the differences among real and synthesized images from different methods.
As shown in Fig.~\ref{fig:combine_seg_user} (b), doctors got high mean accuracy in recognizing the real and synthesized images from cGAN and Sgan (0.64 and 0.55, respectively), but low accuracy (0.43) for PGSgan, which indicates our synthesized images are more realistic. \par

\section{Conclusions}
\label{sec:conclusion}
Here we propose a framework of combining progressive training strategy with guidance auxiliary to synthesize high-resolution US images with high fidelity. Detailed analyses of various experimental results support our assumption that adding background texture information is beneficial for the model to generate realistic US images. In addition, we found that progressive training strategy helps improve performance due to the sharing weights in different phases. Quantitative comparisons, user-study and synthesis edition illustrate the superiority of our proposed method.\par
\bibliographystyle{IEEEbib}
\bibliography{strings}

\end{document}